\documentclass[preprint,aps,nofootinbib,preprintnumbers]{revtex4-1}
\pdfoutput=1

\usepackage{amsmath,amssymb,amsfonts,dcolumn,color,graphicx,graphics,latexsym,placeins,epsfig}
\usepackage{epsfig}
\usepackage{bm}
\usepackage{slashed}
\usepackage{latexsym}
\usepackage{natbib}
\usepackage{url}
\usepackage{dcolumn}
\usepackage{color}
\usepackage{amsfonts,amssymb,amsmath}
\usepackage{graphicx,epsfig}
\usepackage{psfrag}
\usepackage{subfigure}
\usepackage{tabularx}
\usepackage{hyperref}
\hypersetup{colorlinks=true}
\usepackage{comment}

\newcommand{\be}{\begin{equation}}
\newcommand{\ee}{\end{equation}}
\newcommand{\ba}{\begin{eqnarray}}
\newcommand{\ea}{\end{eqnarray}}

\begin{document}

\preprint{IPPP/26/10}
  
\title{Ultralight time-oscillating scalars from magnetized compact stars: electrophilic radiation and photon propagation effects}
\author{Tanmay Kumar Poddar $^{1}$\footnote{tanmay.k.poddar@durham.ac.uk}}
\author{Gaetano Lambiase $^{2}$\footnote{lambiase@sa.infn.it}}
\affiliation{$^{1}$ Institute for Particle Physics Phenomenology (IPPP), Department of Physics, Durham University, Durham DH1 3LE, United Kingdom}
\affiliation{$^{2}$ INFN, Gruppo collegato di Salerno, Via Giovanni Paolo II 132 I-84084 Fisciano (SA), Italy}
\affiliation{$^{2}$ Dipartimento di Fisica E.R. Caianiello, Universit\`a di Salerno,
Via Giovanni Paolo II 132 I-84084 Fisciano (SA), Italy}

\begin{abstract}
Ultralight scalars with electrophilic couplings to the time-dependent Goldreich-Julian charge density of magnetized compact stars can be radiated from their magnetospheres, contributing to pulsar spin-down. Coupling to the time-independent component of the charge density instead generates a quadrupolar scalar field profile, which may influence the orbital dynamics of binary systems. Such scalars can also interact with the time-varying electromagnetic fields of magnetized stars, modifying photon propagation and inducing observable effects in the redshift and residual time-delay measurements, as well as corrections to the background electromagnetic fields. We investigate these phenomena for the Crab pulsar, SGR 1806–20, and GRB 080905A. Using spectral and timing observations, we derive constraints on the scalar-electron and scalar-photon couplings. While the bounds obtained on the scalar-electron coupling from pulsar spin-down are weaker than existing limits, electromagnetic radiation measurements yield the strongest astrophysical constraints to date on the scalar-photon coupling. Compact stars with stronger surface magnetic fields and observations at lower photon frequencies can improve these bounds by several orders of magnitude.
\end{abstract}

\pacs{}
\maketitle

\section{Introduction}

Rotating neutron stars (NSs), including pulsars, provide powerful astrophysical laboratories for probing fundamental physics beyond the Standard Model (SM), such as the nature of dark matter (DM) and properties of the early Universe \cite{Planck:2018vyg,Bramante:2023djs,ParticleDataGroup:2024cfk}. Their extreme densities, strong gravitational and electromagnetic (EM) fields, and macroscopic sizes enable tests of physical regimes that are inaccessible to terrestrial experiments. Depending on their formation history and environment, NSs may exist as isolated objects or as members of binary systems \cite{PortegiesZwart:1997ugk}.

NSs are composed not only of neutrons but also contain substantial populations of electrons, protons, muons, and possibly hyperons in their interiors \cite{Cohen1970}. In addition, NSs typically possess strong dipolar magnetic fields. Under the force-free condition, the magnetosphere corotates with the star and is populated by a dense plasma of charged particles, commonly referred to as the Goldreich-Julian (GJ) magnetosphere \cite{goldreich}.

In general, the rotation axis of a NS is misaligned with its magnetic dipole moment. As a result, the magnetic axis precesses around the spin axis, leading to the emission of EM radiation. The associated loss of rotational energy is dominated by magnetic dipole radiation and manifests as the pulsar's spin-down \cite{Pacini:1967epn,Pacini:1968gfj,Contopoulos:2005rs,Lyne:2010ad,Tauris:2012ex}.

Charged particles in the magnetosphere are accelerated along open magnetic field lines near the magnetic poles, producing synchrotron radiation that is observed as pulsed emission when the radiation beam sweeps across the observer's line of sight. The corresponding spin-down luminosity, which quantifies the rate of rotational energy loss, can be inferred from spectroscopic and timing observations. In addition, the time of arrival of pulsar photons can be measured with extremely high precision \cite{1979ApJ...234.1100D,NANOGrav:2023gor}.

Together, these high-precision timing and spectral measurements provide a sensitive probe of new interactions involving photons and charged particles, enabling stringent constraints on physics beyond the SM.

Key NS properties such as mass, radius, magnetic field strength, and spin frequency as well as binary parameters including the orbital period and semi-major axis, are tightly constrained by multi-wavelength EM observations and, in some cases, gravitational-wave (GW) measurements \cite{Grippa:2024ach}. Consequently, any new interactions involving SM particles within NSs or their magnetospheres can be efficiently probed through precise timing, spectral, and dynamical observations.

In this work, we study the interactions of an ultralight CP-even scalar with electrons in the NS magnetosphere, as well as its dilatonic coupling to photons in pulsar radiation. Such interactions have already been studied in \cite{Poddar:2024thb,Poddar:2025oew}. However, previous studies have typically focused on simplified scenarios involving predominantly time-independent scalar field configurations. In contrast, the present analysis adopts a more general framework, emphasizing time-oscillating scalar field dynamics and systematically exploring the scalar behavior across multiple limiting regimes. This broader treatment significantly enhances the prospects for detection and allows for a more comprehensive assessment of observational signatures.

The scalar field $\varphi$ can interact with the GJ electron number density, $n_{\mathrm{GJ}} \propto \boldsymbol{\Omega}\cdot\mathbf{B}$ \cite{Goldreich:1969sb}, where $\boldsymbol{\Omega}$ denotes the stellar angular velocity and $\mathbf{B}$, the NS magnetic field. Such interactions have been explored previously in the literature \cite{Poddar:2024thb}. However, for simplicity, earlier studies typically assumed that both $\boldsymbol{\Omega}$ and $\mathbf{B}$ possess only azimuthal components, corresponding to a highly restricted field geometry. Under this assumption, the resulting scalar radiation is predominantly dipolar.

In this work, we adopt a more general configuration. We take the rotation axis to be aligned along $\hat{z}$, while allowing the magnetic field to have components along $\hat{r}$, $\hat{\theta}$, and $\hat{\phi}$. This setup describes a rotating, magnetized NS with a misaligned magnetic dipole moment. Within this framework, we compute the scalar radiation emitted by the star and show that the leading contribution is quadrupolar. This follows from the quadrupolar structure of the time-dependent scalar source charge density associated with the GJ magnetosphere. This scalar radiation may contribute to the pulsar spin-down luminosity.

The time-independent component of the scalar source charge density does not give rise to radiation. Instead, it determines the static scalar field configuration outside the NS. We show the resulting profile exhibits a quadrupolar angular structure, together with a Yukawa-type radial dependence. This behavior is more general than the purely radial Yukawa profile obtained in \cite{Poddar:2024thb}, where the angular velocity $\boldsymbol{\Omega}$ and magnetic field $\mathbf{B}$ were assumed to be aligned. In contrast, the present treatment allows for a generic orientation between $\boldsymbol{\Omega}$ and $\mathbf{B}$, leading naturally to a quadrupolar scalar field profile outside the star.

In addition to its interaction with GJ electrons, an ultralight scalar can also couple directly to the time-dependent EM fields of a pulsar. This interaction generates a time-dependent scalar field whose amplitude decreases as $1/r$ at large distances. When photons from pulsar propagate through such a scalar background, their dispersion relation is modified, leading to a change in the group velocity as the photon effectively acquires a scalar-induced mass. As a result, precision measurements of spectral redshift and photon time-of-arrival provide a powerful means to constrain these new interactions.

The vacuum EM fields can be modified through interactions with the scalar field. As a result, the scalar-induced magnetic field is time-oscillating due to the time-dependent behaviors of the scalar field and the background dipolar magnetic field. Precision measurements of EM radiation, and consequently of the inferred surface magnetic field, therefore provide a sensitive probe to constrain the scalar-photon interaction. 

Related effects have been examined previously in the literature \cite{Poddar:2025oew}; however, earlier analyses typically considered time-independent scalar field profiles. In this paper, we consider photon propagation through time-dependent scalar background which are promising for observational tests.

The scalar fields considered in this work are not assumed to be DM. Independent laboratory, astrophysical, and cosmological measurements already impose strong bounds on scalar couplings to photons and electrons. Since our focus is on phenomenological constraints, we do not address scalar mass-generation mechanisms or ultraviolet completions. Ultralight scalars can, however, naturally obtain small masses in a wide class of frameworks e.g., through the clockwork mechanism \cite{Giudice:2016yja,Wood:2023lis}, Planck-suppressed operators in string/quantum-gravity effective theories \cite{Hui:2016ltb,Hubisz:2024hyz,Banerjee:2022wzk}, or non-perturbative effects such as instantons \cite{Kitano:2021fdl}. They also generically arise in extra-dimensional compactifications \cite{Anchordoqui:2023tln} and in scale-invariant or approximate shift-symmetry breaking scenarios \cite{Ferreira:2020fam}.

The paper is organized as follows. In Section.~\ref{scalar-electron}, we study the scalar-electron coupling and the associated electrophilic scalar radiation in pulsar spin-down, and derive the time-independent scalar field profile sourced by the time-independent part of the GJ charge density. In Section.~\ref{scalar-photon}, we compute the time-dependent scalar field profile arising from its interaction with the EM fields of a magnetized star. We further derive the modified photon dispersion relation for radiation propagating through such a scalar background and evaluate the resulting scalar-induced photon redshift and residual time-delay effects. In addition, we obtain the corresponding modifications to the EM fields induced by the scalar interaction. In Section.~\ref{constraint}, we present constraints on the scalar-electron and scalar-photon couplings derived from pulsar spin-down luminosities, as well as from precision measurements of photon redshift, time-of-arrival delays, and EM radiation. Finally, we summarize our results and discuss their implications in Section.~\ref{conclusion}.

We use natural system of units $c$ (speed of light in vacuum) $=\hbar$ (reduced Planck's constant) $=1$ in this paper, unless stated otherwise.

\section{Scalar-electron coupling}\label{scalar-electron}

In this section, we investigate the sourcing of an ultralight scalar field by the GJ electron charge density in the magnetosphere of a rotating, magnetized NS. We analyze the resulting scalar radiation in a realistic geometric configuration and quantify its contribution to the pulsar spin-down. In addition, we derive the time-independent scalar field profile generated by the magnetosphere and discuss its impact on the orbital dynamics of a binary system.

\subsection{Spin-down of magnetized star due to scalar radiation sourced by GJ charge density}\label{spindown}
The EM fields outside a NS is modeled as a rotating, magnetized sphere with a misaligned dipole moment, derived in \cite{1955AnAp...18....1D}. Although the exact vacuum solutions involve spherical Bessel functions of the third kind, the leading-order terms provide a much simpler and widely used description of the magnetospheric fields. Retaining only the lowest multipole order, the time-oscillating magnetic field components take the form
\begin{equation}
\begin{aligned}
B_r &= \frac{ B_0 R^3}{r^3}\left(\cos\alpha \cos\theta+\sin\alpha\sin\theta\cos(\phi-\Omega t)\right),\\
B_\theta &= \frac{B_0 R^3}{2r^3}\left(\cos\alpha\sin\theta-\sin\alpha\cos\theta\cos(\phi-\Omega t)\right),\\
B_\phi &= \frac{B_0 R^3}{2r^3}\sin\alpha\sin(\phi-\Omega t),
\end{aligned}
\label{eq:1}
\end{equation}
and the corresponding time-oscillating electric-field components are
\begin{equation}
\begin{aligned}
E_r &= -\frac{B_0\Omega R^5}{2r^4}\left[\cos\alpha\left(3\cos^2\theta-1\right)
+\frac{3}{2}\sin\alpha\cos(\phi-\Omega t)\sin2\theta\right],\\
E_\theta &= -\frac{B_0 R^3\Omega}{2r^2}\left[\frac{R^2}{r^2}\cos\alpha\sin2\theta
+\sin\alpha\left(1-\frac{R^2}{r^2}\cos2\theta\right)\cos(\phi-\Omega t)\right],\\
  E_\phi &= \frac{B_0 R^3\Omega}{2r^2}\sin\alpha\cos\theta\sin(\phi-\Omega t)
  \left(1-\frac{R^2}{r^2}\right),
  \end{aligned}
  \label{eq:2}
  \end{equation}
where we impose the boundary conditions that the tangential component of the electric field is continuous at $r=R$, while the normal component may be discontinuous across the boundary, $r,\,\theta,\,\phi$ are the standard spherical polar coordinates, $B_0$ stands for the surface magnetic field of the pulsar, $R$ denotes its radius, $\Omega$ is the angular velocity and $\alpha$ is the misalignment angle between the rotation axis and the magnetic moment axis. A non-zero $\alpha$ allows pulsed radiation from the pulsar. In the limit $\alpha\to 0$, corresponds to the aligned rotator model, the magnetic and electric fields from Eqs.~\ref{eq:1} and \ref{eq:2} reduce to
\begin{equation}
\mathbf{B}(r,\theta)=\frac{B_0 R^3}{r^3}\Big(\cos\theta \hat{r}+\frac{\sin\theta}{2}\hat{\theta}\Big), ~~~\mathbf{E}(r,\theta)=-\frac{B_0 \Omega R^5}{2r^4}\Big[(3\cos^2\theta-1)\hat{r}+\sin2\theta\hat{\theta}  \Big]. 
\label{eq:3}
\end{equation}

The electrophilic scalar field $\varphi$ is coupled to the GJ charge density as described by the Lagrangian
\begin{equation}
\mathcal{L}\supset
\frac{1}{2}\partial_\mu\varphi\partial^\mu\varphi
-\frac{1}{2}m_\varphi^{2}\varphi^{2}
-g_e\varphi n^{e}_{\rm GJ},
\end{equation}
where $m_\varphi$ is the scalar mass, $g_e$ denotes scalar's coupling strength to electrons, and the GJ number density is given as
\begin{equation}
n^{e}_{\rm GJ} = -\frac{2\boldsymbol{\Omega}\cdot\mathbf{B}}{e}.
\end{equation}

Therefore, the resulting equation of motion for $\varphi$ becomes
\begin{equation} (\Box+m^2_\varphi )\varphi=\frac{2g_e\mathbf{\Omega}\cdot \mathbf{B}}{e}. \label{eq:5} 
\end{equation}

Considering the stellar rotation axis to align with $\hat{z}$, one may write
\begin{equation}
\boldsymbol{\Omega}=\Omega\hat{z}
= \Omega(\cos\theta\hat{r}-\sin\theta\hat{\theta}),
\end{equation}
which leads to the scalar field equation of motion from Eq.~\ref{eq:5} as
\begin{equation} 
(\Box+m^2_\varphi )\varphi= \frac{g_e B_0 R^3\Omega}{er^3} \cos\alpha(3\cos^2\theta-1)+\frac{3}{2}\frac{g_e B_0R^3\Omega}{er^3}\sin\alpha\sin 2\theta\cos(\phi-\Omega t).
\label{eq:6}
\end{equation}

The first term is stationary and therefore does not generate any time-varying scalar radiation. By contrast, the second term oscillates as $\cos(\phi-\Omega t)$ and constitutes the source of pulsed scalar emission, ultimately contributing to the pulsar's spin-down power. Therefore, the relevant scalar-field equation of motion for pulsar spin-down is
\begin{equation}
(\Box+m^2_\varphi )\varphi= \frac{3}{2}\frac{g_e B_0R^3\Omega}{er^3}\sin\alpha\sin 2\theta\cos(\phi-\Omega t).    
\label{eq:7}
\end{equation}

The scalar-induced source charge density can be expressed as
\begin{equation}
\rho_{\Omega}(\mathbf{r})
= S_{0}\frac{\sin\alpha}{r^{3}}
\sin 2\theta e^{i\phi},
\qquad
S_{0} = \frac{3}{2}\frac{g_{e} B_{0} R^{3}\Omega}{e}.
\label{eq:8}
\end{equation}
The angular structure of $\rho_{\Omega}(\mathbf{r})$ may be recast in terms of spherical harmonics as
\begin{equation}
\rho_{\Omega}(\mathbf{r})
= C_{2}\frac{Y_{2,1}(\theta,\phi)}{r^{3}},
\qquad
C_{2}
= -\frac{3 g_{e} B_{0} R^{3}\Omega}{e}
\sin\alpha\sqrt{\frac{8\pi}{15}},~~~Y_{2,1}(\theta,\phi)
= -\sqrt{\frac{15}{8\pi}}
\sin\theta\cos\theta e^{i\phi}.
\label{eq:9}
\end{equation}
Thus, the source corresponds to a purely quadrupolar configuration with $(\ell,m)=(2,\pm1)$. The $m=-1$ component arises from the complex-conjugate mode.

The corresponding multipole moment is obtained from
\begin{equation}
Q_{2,1}
= \int d^{3}r
\rho_{\Omega}(\mathbf{r})r^2
Y^{*}_{2,1}(\hat r)
\simeq \frac{C_{2}R_{\rm LC}^{2}}{2},
\label{eq:10}
\end{equation}

where the radial integration is taken over $R \le r \le R_{\rm LC}$, where $R_{\rm LC}$ denotes the light cylinder (LC) radius, and the final expression follows from $R_{\rm LC} \gg R$. An identical magnitude is obtained for $Q_{2,-1}$.

The outgoing scalar wave solution at frequency $\Omega$ in the far-field limit is
\begin{equation}
\varphi_{\mathrm{\Omega}}(r, \hat{n})=\frac{e^{ik\cdot r}}{r}\sum_{l,m} i^l\frac{k^l}{(2l+1)!!}Y_{l,m}(\hat{n})Q_{l,m},  \label{eq:11}  
\end{equation}
with $\hat{n}$ denotes the unit vector specifying the observation direction and $k=\sqrt{\Omega^2-m^2_\varphi}$. Therefore, the time-averaged power carried by a real scalar at frequency $\Omega$ is \cite{PhysRevD.49.6892}
\begin{equation}
P_\Omega=\frac{\Omega k}{2}\sum _{l,m}\Big(\frac{k^l}{(2l+1)!!}\Big)^2|Q_{l,m}|^2.  
\label{eq:12}  
\end{equation}
Thus, for $(l,m)=(2,\pm 1)$, the total power loss due to such scalar radiation is
\begin{equation}
P_\Omega \simeq \frac{2\pi}{375e^2}g^2_e B^2_0 R^6\Omega^4\sin^2\alpha\Big(1-\frac{m^2_\varphi}{\Omega^2}\Big)^{5/2},
\label{eq:13}
\end{equation}
where we use $R_{\mathrm{LC}}\sim 1/\Omega$. The scalar radiation is only possible for $m_\varphi\lesssim\Omega$, obtained from the kinematics.

\subsection{Time-independent scalar field profile sourced by the GJ magnetosphere} \label{field-profile}

In this section, we derive the full analytic time-independent profile of a massive scalar field that couples to the GJ charge density in the magnetosphere of a rotating NS. The coupling induces an effective scalar source that is confined to the region between the stellar surface and the LC. The scalar field satisfies the massive Klein-Gordon equation,
\begin{equation}
(\nabla^{2}-m_\varphi^{2})\varphi(r,\theta)=\mathcal{S}(r,\theta).
\label{eq:KG_main}
\end{equation}

The GJ density induces a quadrupolar time-independent source of the form given by the first term of Eq.~\ref{eq:6} as
\begin{equation}
\mathcal{S}(r,\theta)=
-\frac{2 \mathcal{S}_0}{r^{3}}P_{2}(\cos\theta),
\qquad
R< r < R_{\rm LC},
\label{eq:source}
\end{equation}
where 
\begin{equation}
\mathcal{S}_0=\frac{g_e B_0 R^3\Omega}{e}\cos\alpha, \qquad P_2(\cos\theta)=\frac{1}{2}(3\cos^2\theta-1),    
\end{equation}
and $\mathcal{S}_0=0$ for $r<R$ and $r>R_\mathrm{LC}$. The time-dependence in the source charge density and hence in the scalar field profile is important in the study of scalar radiation which we already discuss in Section~\ref{spindown}. We now analyze the contribution from the time-independent component of the source charge density. This term gives rise to a non-gravitational long-range interaction that can alter the orbital dynamics of a binary system.

We therefore expand the scalar field as
\begin{equation}
\varphi(r,\theta)=f(r)P_{2}(\cos\theta),
\label{eq:separation}
\end{equation}
which reduces Eq.~\ref{eq:KG_main} to the radial equation
\begin{equation}
f''(r)+\frac{2}{r}f'(r)-\frac{6}{r^{2}}f(r)-m_\varphi^{2}f(r)
=-\frac{2\mathcal{S}_0}{r^{3}},
\qquad R<r<R_{\rm LC},
\label{eq:radial_source}
\end{equation}
and to the homogeneous version of this equation in the regions $r<R$ and $r>R_{\mathrm{LC}}$.

The boundary conditions are: $f(r)$ must be finite as $r\to 0$, $f(r)$ and $f^\prime (r)$ are continuous at $r=R$ and $r=R_{\mathrm{LC}}$ and the radial function $f(r)$ must go to zero as $r\to \infty$. In addition, we introduce dimensionless variables $x=m_\varphi r,~ x_{\mathrm{NS}}=m_\varphi R,~x_{\mathrm{LC}}=m_\varphi R_{\mathrm{LC}}$ for convenience. Therefore, the homogeneous part of Eq.~\ref{eq:radial_source} admits the standard modified spherical Bessel functions of order $l=2$ as
\begin{equation}
i_2(x)=\frac{(x^2+3)\sinh x-3x\cosh x}{x^3},~~~k_2(x)=\frac{e^{-x}}{x}\Big(1+\frac{3}{x}+\frac{3}{x^2}\Big),
\label{nwq}
\end{equation}
where $i_2(x)$ is regular at the origin and $k_2(x)$ decays exponentially at large distances, from the boundary conditions.

Thus, the Green's function solution for the $l=2$ mode is
\begin{equation}
f(r)=2S_{0}m_\varphi\int_{R}^{R_{\rm LC}}
\frac{dr^\prime}{r^\prime}i_{2}(m_\varphi r_<)k_{2}(m_\varphi r_>),
\label{eq:green_basic}
\end{equation}
where $r_<=\mathrm{min}(r,r^\prime)$ and $r_>=\mathrm{max}(r,r^\prime)$. This expression simultaneously enforces regularity at the center, continuity across both boundaries, and decay at infinity.

For the scalar field profile outside the LC $(r>R_{\mathrm{LC}})$, we obtain the field profile from the Green's function method as 
\begin{equation}
\begin{split}
\varphi_{\mathrm{out}}(r, \theta)=2\mathcal{S}_0m_\varphi\Big[\Big(\frac{\cosh x_{\mathrm{LC}}}{x^2_{\mathrm{LC}}}-\frac{\sinh x_{\mathrm{LC}}}{x^3_{\mathrm{LC}}}\Big)-\Big(\frac{\cosh x_{\mathrm{NS}}}{x^2_{\mathrm{NS}}}-\frac{\sinh x_{\mathrm{NS}}}{x^3_{\mathrm{NS}}}\Big)\Big]k_2(x)
P_2(\cos\theta),
\label{outside}
\end{split}
\end{equation}
and for the scalar field profile inside the LC $(R<r<R_{\mathrm{LC}})$, 
we obtain
\begin{equation}
\begin{split}
\varphi_{\mathrm{int}}(r,\theta)=2\mathcal{S}_0 m_\varphi \Big\{k_2(x)\Big[\Big(\frac{\cosh x}{x^2}-\frac{\sinh x}{x^3}\Big)-\Big(\frac{\cosh x_{\mathrm{NS}}}{x^2_\mathrm{NS}}-\frac{\sinh x_{\mathrm{NS}}}{x^3_{\mathrm{NS}}}\Big)\Big]+\\
i_2(x)\Big[\frac{(x+1)e^{-x}}{x^3}-\frac{(x_{\mathrm{LC}}+1)e^{-x_{\mathrm{LC}}}}{x^3_{\mathrm{LC}}}\Big]\Big\}P_2(\cos\theta). 
\end{split}
\label{inside}
\end{equation}
In the massless scalar limit, $x,~x_{\mathrm{NS}},~x_{\mathrm{LC}}\to 0$, we obtain the scalar field profile by expanding $i_2$, $k_2$, $\cosh$, and $\sinh$ from  Eqs.~\ref{outside} and Eq.~\ref{inside} as
\begin{equation}
\varphi_{\mathrm{out}}(r,\theta)\to \frac{\mathcal{S}_0}{5}\frac{R^2_{\mathrm{LC}}}{r^3} P_2(\cos\theta),~~~  \varphi_{\mathrm{int}}(r,\theta)\to \Big[-\frac{2\mathcal{S}_0}{15}\frac{r^2}{R^3_{\mathrm{LC}}}+\frac{\mathcal{S}_0}{3r}-\frac{\mathcal{S}_0 R^2}{5r^3}\Big]P_2(\cos\theta).
\end{equation}
In the massive far-zone limit $m_\varphi r\gg 1$ and $r\gg R_{\mathrm{LC}}$, $k_2(x)\sim e^{-x}/x$ and Eq.~\ref{outside} reduces to the Yukawa tail given as
\begin{equation}
\varphi_{\mathrm{out}}(r,\theta)\simeq \mathcal{Q}_2\frac{e^{-m_\varphi r}}{r} P_2(\cos\theta),~~~\mathcal{Q}_2=2\mathcal{S}_0\Big[\Big(\frac{\cosh x_{\mathrm{LC}}}{x^2_{\mathrm{LC}}}-\frac{\sinh x_{\mathrm{LC}}}{x^3_{\mathrm{LC}}}\Big)-\Big(\frac{\cosh x_{\mathrm{NS}}}{x^2_{\mathrm{NS}}}-\frac{\sinh x_{\mathrm{NS}}}{x^3_{\mathrm{NS}}}\Big)\Big], 
\label{oneq}
\end{equation}
where $\mathcal{Q}_2$ is the effective quadrupole moment. If the scalar field is light over the magnetosphere, then $m_\varphi R\ll1$ and $m_\varphi R_{\mathrm{LC}}\ll1$ and using Eq.~\ref{oneq} we obtain the far field solution as
\begin{equation}
\varphi_{\mathrm{out}}^\mathrm{eff}(r,\theta)\approx\frac{g_eB_0R^3\Omega m^2_\varphi R^2_{\mathrm{LC}}}{30e}\Big(\frac{e^{-m_\varphi r}}{r}\Big)\cos\alpha(3\cos^2\theta-1),\qquad m_\varphi\lesssim 1/R_{\mathrm{LC}},
\label{finalphi}
\end{equation}
where we consider $R_{\mathrm{LC}}\gg R$. Therefore, the scalar filed has a Yukawa type quadrupolar behavior. 

The exterior scalar field solution in Eq.~\ref{finalphi}, $\varphi_{\rm out}^{\rm eff}(r,\theta) \propto (e^{-m_\varphi r}/r) P_2(\cos\theta)$, implies that each star supports only quadrupolar scalar hair. For a binary companion with the same angular structure, no leading monopole-monopole fifth force is generated. The dominant scalar-mediated interaction instead arises from quadrupole coupling to the tidal field, modulated by the quadrupolar angular factor. This contribution is parametrically suppressed compared to gravity, rendering it a sub-leading and significantly weaker correction to the Newtonian gravitational force that governs the orbital dynamics.

\section{Scalar-photon coupling}\label{scalar-photon}
We next examine the interaction between an ultralight scalar field and EM radiation emitted by magnetized NSs. We derive the resulting time-oscillating scalar field profile generated through its coupling to the stellar EM field, and determine the corresponding modifications to photon propagation. In particular, we obtain the scalar-induced corrections to the photon dispersion relation and the associated modifications of the EM fields. We further compute the resulting photon residual time delay arising from the effective scalar-induced photon mass.

\subsection{Photon propagation in a time-oscillating scalar field background around magnetized stars}\label{redshift}

The CP even scalar may also interact with the photon through the interaction Lagrangian 
\begin{equation}
\mathcal{L}=\frac{1}{2}g_{\varphi\gamma\gamma}\varphi F_{\mu\nu}F^{\mu\nu},
\end{equation}
with the scalar-induced source charge density given as $\rho_\varphi=g_{\varphi\gamma\gamma}(B^2-E^2)$, since $F_{\mu\nu}F^{\mu\nu}/2=(B^2-E^2)$. Using Eqs.~\ref{eq:1} and \ref{eq:2} we write
\begin{equation}
\rho_\varphi\approx\frac{g_{\varphi\gamma\gamma}B^2_0 R^6}{r^6}\Big[\frac{3}{2}\sin\alpha\cos\alpha\sin\theta\cos\theta \cos(\phi-\Omega t)+\frac{3}{4}\sin^2\alpha\sin^2\theta\cos^2(\phi-\Omega t)\Big], 
\end{equation}

where we retain only the oscillatory components that contribute to scalar radiation and consider $B^2\gg E^2$ as $\Omega R\ll 1$. We note that the source charge density contains harmonics at $\Omega$ and $2\Omega$. Accordingly, we compute the far-zone time-dependent scalar field at these two frequencies.

Therefore, we can write the scalar-induced source charge densities at the two frequencies in terms of spherical harmonics as
\begin{equation}
\rho_{+\Omega}(\mathbf r)
= \frac{3}{2}g_{\varphi\gamma\gamma}\mu^2\sin(2\alpha)
\sqrt{\frac{8\pi}{15}}
\frac{Y_{2,1}(\theta,\phi)}{r^6}, ~~~\rho_{+2\Omega}(\mathbf r)
= -\frac{3}{4}g_{\varphi\gamma\gamma}\mu^2\sin^2\alpha
\sqrt{\frac{32\pi}{15}}
\frac{Y_{2,2}(\theta,\phi)}{r^6},
\label{aksource}
\end{equation}

with $\mu = B_0R^3/2$ and $Y_{l,m}(\theta,\phi)$ are spherical harmonics. These are the spatial parts of the sources oscillating as $e^{-i\Omega t}$ and $e^{-2i\Omega t}$, respectively.

For each frequency $\omega$, we have the equation of motion of the scalar field as
\begin{equation}
(\nabla^2 + k^2)\varphi_\omega(\mathbf r) = -\rho_\omega(\mathbf r),
\qquad k = \sqrt{\omega^2 - m_\varphi^2}.
\end{equation}

In the far-field and long-wavelength limit, the scalar field profile becomes
\begin{equation}
\varphi_\omega(\mathbf r) \simeq
\frac{e^{ikr}}{r}
\sum_{l,m}
i^l\frac{k^l}{(2l+1)!!}
Y_{lm}(\hat{\mathbf r})
Q_{lm}(\omega),
\label{profile1}
\end{equation}
with multipole moments
\begin{equation}
Q_{lm}(\omega)
= \int d^3r^\prime\rho_\omega(\mathbf r^\prime){r^\prime}^l Y^*_{lm}(\hat{\mathbf r}^\prime).
\end{equation}
The sources in Eq.~\ref{aksource} are pure $l=2$, so only $l=2$ terms contribute. Therefore, the relevant quadrupole moments are
\begin{equation}
Q_{2,1}(\Omega)
=\frac{3}{2}g_{\varphi\gamma\gamma}\mu^2\sin(2\alpha)
\sqrt{\frac{8\pi}{15}}\frac{1}{R},~~~Q_{2,2}(2\Omega)
=-\frac{3}{4}g_{\varphi\gamma\gamma}\mu^2\sin^2\alpha
\sqrt{\frac{32\pi}{15}}\frac{1}{R}.
\end{equation}

Using Eq.~\ref{profile1}, we obtain
\begin{equation}
\varphi_\Omega(\mathbf r)
\simeq
-\frac{g_{\varphi\gamma\gamma}B_0^{,2}R^{5}}{40}
\sqrt{\frac{8\pi}{15}}
k_\Omega^{2}
\frac{e^{ik_\Omega r}}{r}
\sin(2\alpha)
Y_{2,1}(\theta,\phi),
\end{equation}
and the time-oscillating scalar field at frequency $\Omega$ becomes
\begin{equation}
\varphi_\Omega(t,r,\theta,\phi)
= \Re\big[\varphi_\Omega(\mathbf r)e^{-i\Omega t}\big]
= -\frac{g_{\varphi\gamma\gamma}B_0^{,2}R^{5}}{40}
\sqrt{\frac{8\pi}{15}}
k_\Omega^{2}
\frac{1}{r}
\sin(2\alpha)
Y_{2,1}(\theta,\phi)
\cos\big(k_\Omega r-\Omega t\big).
\end{equation}
Similarly, the time-oscillating scalar field at frequency $2\Omega$ becomes
\begin{equation}
\varphi_{2\Omega}(t,r,\theta,\phi)
\frac{g_{\varphi\gamma\gamma}B_0^{,2}R^{5}}{80}
\sqrt{\frac{32\pi}{15}}
k_{2\Omega}^{2}
\frac{1}{r}
\sin^2\alpha,
Y_{2,2}(\theta,\phi)
\cos\big(k_{2\Omega} r-2\Omega t\big),
\end{equation}
with
\begin{equation}
k_\Omega=\sqrt{\Omega^2-m^2_\varphi}, \qquad k_{2\Omega}=\sqrt{4\Omega^2-m^2_\varphi}.    
\end{equation}
Thus, the total time-dependent scalar field profile at far zone can be written as
\begin{equation}
\begin{split}
\varphi(t,r,\theta,\phi)
\simeq
-\frac{g_{\varphi\gamma\gamma}B_0^2R^5}{40}
k_\Omega^2\frac{1}{r}
\sin(2\alpha)
\sin\theta\cos\theta
\cos\big(k_\Omega r - \Omega t + \phi\big)
+\\
\frac{g_{\varphi\gamma\gamma}B_0^2R^5}{80}
k_{2\Omega}^2\frac{1}{r}
\sin^2\alpha
\sin^2\theta
\cos\big(k_{2\Omega} r - 2\Omega t +2\phi\big),
\end{split}
\label{masterphoton}
\end{equation}
and we are working in the far-field, long-wavelength limit as $r\gg R, k_\Omega R\ll1, k_{2\Omega}R\ll1$. Hence, the scalar field is time-oscillating quadrupolar in nature.

As EM radiation propagates through a long-range time-oscillating scalar-field background, the Maxwell equations governing its dynamics are modified as\cite{Domcke:2023bat}
\begin{eqnarray}
\nabla\cdot \mathbf{E}&=&-g_{\varphi\gamma\gamma}\mathbf{E}\cdot \nabla\varphi,\nonumber\\
\nabla\times \mathbf{B}&=&\frac{\partial \mathbf{E}}{\partial t}-g_{\varphi\gamma\gamma}\nabla\varphi \times \mathbf{B}+g_{\varphi\gamma\gamma}\frac{\partial \varphi}{\partial t}\mathbf{E},\nonumber\\
\nabla\cdot \mathbf{B}&=&0,\nonumber\\
\nabla\times \mathbf{E}&=&-\frac{\partial \mathbf{B}}{\partial t},
\label{ewq}
\end{eqnarray}
in the absence of any source plasma charge and current densities. Neglecting terms which appear as two spatial and temporal derivatives of $\varphi$, the wave equation for $\mathbf{B}$ from Eq.~\ref{ewq} can be written as
\begin{equation}
\Box \mathbf{B}\approx g_{\varphi\gamma\gamma}(\nabla \varphi\cdot \nabla)\mathbf{B}-g_{\varphi\gamma\gamma}\dot{\varphi}\dot{\mathbf{B}},
\label{combinedwave}
\end{equation}
where we consider the photon propagation is only radial. We choose the Eikonal ansatz $\mathbf{B}(x,t)=\mathcal{B}e^{i S(x,t)}$, where the phase $S(x,t)$ measures the frequency $\omega$ and wavenumber $k$ of the photon as $\omega=-\partial S/\partial t$ and $k=\nabla S$. 

Therefore, we obtain the dispersion relation from Eq.~\ref{combinedwave} as
\begin{equation}
\omega^2=k^2-ig_{\varphi\gamma\gamma}(k\cdot \nabla \varphi+\dot{\varphi}\omega).
\label{dispersion}
\end{equation}
Solving Eq.~\ref{dispersion} for $k$, we obtain
\begin{equation}
k=k_R+ik_I\simeq \omega\Big(1-\frac{m^2_{\gamma_k}}{8\omega^2}\Big) +\frac{i}{2}(m_{\gamma_k}+m_{\gamma_\omega}),  
\label{sol-k}
\end{equation}
where $k_R=\omega-(m^2_{\gamma_k}/8\omega)$, $k_I=(m_{\gamma_k}+m_{\gamma_\omega})/2$, with the definitions of the scalar-induced photon mass $m_{\gamma_k}=|g_{\varphi\gamma\gamma}\hat{\bm k}\cdot \nabla\varphi|$ and $m_{\gamma_\omega}=|g_{\varphi\gamma\gamma}\dot{\varphi}|$. We write the explicit expressions of $m_{\gamma_k}$ and $m_{\gamma_\omega}$ as
\begin{align}
\label{import}
m_{\gamma_k}&\approx\frac{g^2_{\varphi\gamma\gamma}B^2_0 R^5\Omega^2}{80 r^2}\Big(1-\frac{m^2_\varphi}{\Omega^2}\Big)\sin(2\alpha)\sin(2\theta)\Big[\cos(k_\Omega r-\Omega t+\phi)+k_\Omega r\sin(k_\Omega r-\Omega t+\phi)\Big],\\
m_{\gamma_\omega}&\approx \frac{g^2_{\varphi\gamma\gamma}B^2_0R^5\Omega^2}{80 r^2}\Big(1-\frac{m^2_\varphi}{\Omega^2}\Big)\sin(2\alpha)\sin(2\theta)(\Omega r)\sin(k_\Omega r-\Omega t+\phi),
\end{align}
where we consider $\hat{\bm k}$ and $\nabla\varphi$ are parallel along the radial direction. Therefore, in the limit $\Omega r\ll1$, $m_{\gamma_k}\gg m_{\gamma_\omega}$ and we write the group velocity $v_g$ of photon from Eq.~\ref{dispersion} as
\begin{equation}
v_g\approx 1-\frac{m^2_{\gamma_k}}{8\omega^2}\approx 1-\frac{g^4_{\varphi\gamma\gamma}B^4_0R^{10}\Omega^4}{51200\omega^2 r^4}\Big(1-\frac{m^2_\varphi}{\Omega^2}\Big)^2\sin^2(2\alpha)\sin^2(2\theta)\cos^2(k_\Omega r-\Omega t+\phi).    
\end{equation}
Thus, near the surface of the star $(r\sim R)$, the photon group velocity can be written as
\begin{equation}
v_g\approx 1-\frac{g^4_{\varphi\gamma\gamma}B^4_0R^6\Omega^4}{51200\omega^2}\Big(1-\frac{m^2_\varphi}{\Omega^2}\Big)^2\sin^2(2\alpha)\sin^2(2\theta)\cos^2(k_\Omega R-\Omega t+\phi).    
\end{equation}
The scalar-photon interaction alters the photon group velocity only at order $\mathcal{O}(g^4_{\varphi\gamma\gamma})$. Consequently, in the limit $g_{\varphi\gamma\gamma}\to 0$, the correction disappears, and the group velocity reduces to its standard vacuum value $v_g\to 1$.

The modification of the photon dispersion relation induced by the scalar background leads to a shift in the apparent redshift of a photon with wavelength $\lambda$ as measured in asymptotically flat spacetime. Since only the real part of the wavenumber, $k_R$, contributes to the observed redshift, the change in redshift between the emission point $r_1 = R$ and a distant observer at $r_2 \to \infty$ is

\begin{equation}
\begin{split}
\delta z (t)
= \frac{\lambda(r_2)-\lambda(r_1)}{\lambda(r_1)}
= \frac{k_R(r_1)-k_R(r_2)}{k_R(r_2)}
\simeq \frac{m_{\gamma_k}^2}{8\omega^2}\approx \\ \approx\frac{g^4_{\varphi\gamma\gamma}B^4_0R^6\Omega^4}{102400\omega^2}\Big(1-\frac{m^2_\varphi}{\Omega^2}\Big)^2\sin^2(2\alpha)
\sin^2(2\theta)(1+\cos(2k_\Omega R-2\Omega t+2\phi)).
\end{split}
\label{redshift-master}
\end{equation}
 The redshift modification becomes significant when $m_{\gamma_k}$ is comparable to the photon frequency $\omega$, favoring systems with strong surface magnetic fields, rapid rotation, and observations at low photon frequencies. Since the induced redshift varies periodically in time rather than being static, it exhibits characteristic features that help disentangle it from conventional redshift contributions. Moreover, such a modulation is possible only when the scalar mass satisfies $m_\varphi<\Omega$.

Equation~\ref{redshift-master} describes a time-dependent redshift modulation imprinted on photons at emission and persisting coherently to detection. The oscillatory component of $\delta z(t)$ is a sinusoid at angular frequency $2\Omega$, superimposed on a constant (DC) offset. This fractional frequency shift maps directly onto pulsar timing residuals, defined as the difference between the observed and model predicted times of arrivals (TOAs). The signal can therefore be isolated by performing a coherent template search in the residuals, using a targeted matched-filter or sinusoidal fit at frequency $2\Omega$, enabling sensitivity to extremely small, phase-coherent periodic effects.

The scalar-induced redshift contribution in Eq.~\ref{redshift-master} can be cleanly separated from standard backgrounds. Unlike gravitational redshift, which is constant and frequency-independent, the scalar signal yields a coherent, time-modulated fractional frequency shift locked to the second rotational harmonic, $2\Omega$, and scales as $\omega^{-2}$. While plasma effects share the same $\omega^{-2}$ scaling, they generate Faraday rotation--an effect that is absent for the scalar-induced modulation. In addition, interstellar dispersion-measure delays are neither phase-locked nor temporally coherent at $2\Omega$. By contrast, the scalar signal remains strictly phase-coherent and ephemeris-locked to $2\Omega$, providing a distinctive clock-like imprint that can be optimally extracted through matched-filter or sinusoidal template searches in pulsar timing residuals.

The imaginary part of the wavenumber contributes to the photon absorption in the scalar field background. If a photon with initial intensity $I_0$ propagates through an absorptive scalar field background of propagation length $x$, then the resultant intensity becomes $I=I_0 e^{-\alpha x}$, where $\alpha$ represents the absorption coefficient given as
\begin{equation}
\alpha=2k_I=m_{\gamma_k}+m_{\gamma_\omega}=\frac{1}{x}\ln\Big(\frac{I_0}{I}\Big).  
\end{equation}
There can also be exponential enhancement in the field amplitude and hence in the photon intensity if $\bm{\hat{k}}$ and $\nabla\phi$ are anti-parallel to each other. However, the modifications in the group velocity and change in photon redshift would remain same as they both appear in even powers of $m_{\gamma_k}$.
\subsection{Scalar-induced EM fields}
The interaction of a CP-even scalar field with the EM fields of the star leads to modified Maxwell's EM equations in vacuum. We treat these modifications perturbatively by expanding the EM field strength tensor in powers of the scalar-photon coupling $g_{\phi\gamma\gamma}$,
\begin{equation}
F^{\mu\nu}=F^{\mu\nu}_{(0)}+F^{\mu\nu}_\varphi +\mathcal{O}(g^2_{\varphi\gamma\gamma}),
\label{sem1}
\end{equation}
where the subscript `$(0)$' denotes quantities evaluated in the limit $g_{\varphi\gamma\gamma}=0$. We keep the leading correction term, which is linear in $g_{\varphi\gamma\gamma}$ and obtain the equation of motion for $F^\varphi_{\mu\nu}$ as
\begin{equation}
\partial_\mu F^{\mu\nu}_{\varphi}
=-g_{\varphi\gamma\gamma}(\partial_\mu\varphi)F^{\mu\nu}_{(0)},
\end{equation}
in the absence of plasma charge and current densities. Together with the Bianchi identity, this relation determines the scalar-induced electric and magnetic fields, $\mathbf{E}_\varphi$ and $\mathbf{B}_\varphi$, in terms of the background fields $\mathbf{E}_{(0)}$ and $\mathbf{B}_{(0)}$, given in Eqs.~\ref{eq:1} and \ref{eq:2}.

Following section.~\ref{redshift}, we obtain the wave equation governing the scalar-induced magnetic field,
\begin{equation}
\Box \mathbf{B}_\varphi\approx g_{\varphi\gamma\gamma}(\nabla\varphi\cdot \nabla)\mathbf{B}_{(0)}-g_{\varphi\gamma\gamma}\dot{\varphi}\dot{\mathbf{B}}_{(0)}.   
\label{sem2}
\end{equation}

Working in the regime $\Omega r \ll 1$, where spatial gradients dominate over temporal variations $(\nabla\varphi \gg \dot{\varphi})$, the scalar-induced magnetic field obeys
\begin{equation}
\Box \mathbf{B}_\varphi \simeq g_{\varphi\gamma\gamma}(\nabla\varphi\cdot\nabla)\mathbf{B}_{(0)} .
\label{sem3}
\end{equation}
Under this approximation, the wave equation for the radial component of the scalar-induced magnetic field reduces to
\begin{equation}
\begin{split}
\Box B^r_\varphi
= -\frac{3}{80} g_{\varphi\gamma\gamma}^2 B_0^3 R^8 \Omega^2
\Big(1-\frac{m_\varphi^2}{\Omega^2}\Big)\sin(2\alpha)
\frac{1}{r^6}
\Big[
\cos\alpha \cos\theta \sin2\theta \cos(\phi-\Omega t)+
\\
\sin\alpha \sin\theta \sin2\theta \cos^2(\phi-\Omega t)
\Big],
\end{split}
\label{sem4}
\end{equation}
where Eqs.~\ref{eq:1} and \ref{masterphoton} have been used and terms suppressed by $\Omega r\ll1$ have been neglected. We also assume $\nabla\varphi$ is radial for simplicity. In the limit $\Omega r\ll1$, $\partial_t^2B_r\ll \nabla^2B_r$ and the time-derivative term in the d'Alembertian can be consistently dropped, which results Eq.~\ref{sem4} to a three-dimensional Poisson equation. 

Imposing the boundary conditions that $B^r_\varphi\to0$ as $r\to\infty$ and that $B^r_\varphi$ remains finite at the stellar surface $r=R$, we obtain the leading-order exterior solution
\begin{equation}
B^r_\varphi \simeq
\frac{3}{2000} g_{\varphi\gamma\gamma}^2 B_0^3 R^8 \Omega^2
\Big(1-\frac{m_\varphi^2}{\Omega^2}\Big)\sin(2\alpha)
\frac{\cos\alpha}{r^4}
\sin\theta\cos(\phi-\Omega t),
\label{sem5}
\end{equation}
which exhibits a dipolar angular structure.

In addition to this leading $l=1$ contribution, an $l=3$ component is also generated through angular momentum mixing, characterized by a logarithmic radial dependence $\sim r^{-4}\ln(r/R)$. However, this octupolar term is subdominant near the stellar surface and is therefore neglected in the present analysis. Since the background magnetic field is dipolar $(l=1)$ and the scalar field sourced by the rotating magnetosphere is quadrupolar $(l=2)$, their coupling naturally produces scalar-induced magnetic field components with $l=1$ and $l=3$.

Furthermore, time-independent contributions, as well as higher harmonics at frequency $2\Omega$, do not contribute (or subdominant) to radiative or modulation-based observables and are therefore omitted. The scalar-induced magnetic field falls off as $r^{-4}$, in contrast to the $r^{-3}$ behavior of the background dipolar field, and is proportional to $g_{\phi\gamma\gamma}^2$, making it parametrically suppressed. Consequently, the modification of the surface magnetic field due to the scalar interaction is small. The induced magnetic field exhibits a time modulation at the stellar rotation frequency $\Omega$, and the effect is relevant only for scalar masses satisfying $m_\varphi \lesssim \Omega$.

In a similar fashion, we may calculate $B_\varphi^\theta$ and $B_\varphi^\phi$ and scalar-induced electric field. However, to set a conservative bound on $g_{\phi\gamma\gamma}$ from the estimation of surface magnetic field, only $B^r_\varphi$ is sufficient.

\subsection{Scalar-induced photon mass and the residual/dedispersed time delay}\label{dedispersed}

In Section.~\ref{redshift}, we derived the modification of the photon dispersion relation induced by the scalar-photon interaction, which effectively endows the photon with a small mass. Here we focus on the resulting frequency-dependent propagation effect and compute the relative arrival-time delay between photons of different frequencies emitted from the same astrophysical source. For massive photons, signals at different frequencies propagate with different group velocities, leading to an observable time delay over astrophysical distances. Expanding the scalar-modified dispersion relation to $\mathcal{O}(\nu^{-4})$, the propagation-induced time delay between two frequencies $\nu_1$ and $\nu_2>\nu_1$ from a source at distance $d$ can be written as \cite{Chang:2022qct}
\begin{equation}
\Delta t_{\rm 1st}^\varphi(\nu^{-2})+\Delta t^\varphi_{\rm 2nd}(\nu^{-4})\simeq \int dl\Big[\frac{\xi}{2}\Big(\frac{1}{\nu_1^2}-\frac{1}{\nu_2^2}\Big)+\frac{3\xi^2}{8}\Big(\frac{1}{\nu_1^4}-\frac{1}{\nu_2^4}\Big)\Big],   
\label{ddis1}
\end{equation}
where $\xi \equiv m_{\gamma_k}^2/(16\pi^2)$. The term $\mathcal{O}(\nu^{-4})$ is retained because standard dedispersion procedures remove contributions proportional to $1/\nu^2$, rendering the next-to-leading term potentially observable. For nearby sources, cosmological redshift effects can be neglected, and the line-of-sight integral reduces to the source distance $d$. The time delay then simplifies to
\begin{equation}
\Delta t_{\rm 1st}^\varphi+\Delta t^\varphi_{\rm 2nd}\simeq \frac{d~\xi}{2}\Big(\frac{1}{\nu^2_1}-\frac{1}{\nu^2_2}\Big)+\frac{3d\xi^2}{8}\Big(\frac{1}{\nu^4_1}-\frac{1}{\nu^4_2}\Big).   
\label{ddis2}
\end{equation}

Therefore, the total observed time delay between two photons of different frequencies can be written as
\begin{equation}
\Delta t_{\rm obs}=\Delta t_{\rm int}+\Delta t^{\rm \mathcal{DM}}_{\rm 1st} +\Delta t^{\rm \mathcal{DM}}_{\rm 2nd}+\Delta t_{\rm 1st}^\varphi+\Delta t^\varphi_{\rm 2nd},
\label{ddis3}
\end{equation}
where $\Delta t_{\rm int}$ denotes the intrinsic emission delay associated with the radiation mechanism and source geometry, and $\Delta t^{\rm\mathcal{DM}}$ represents the plasma-induced dispersion delay referring dispersion measure ($\mathcal{DM}$). Standard de-dispersion procedures remove all contributions proportional to $1/\nu^{2}$. Consequently, after de-dispersion, the residual time delay is given by
\begin{equation}
\Delta t_{\rm res}=  \Delta t_{\rm obs}-\Delta t^{\rm \mathcal{DM}}_{\rm 1st}-\Delta t_{\rm 1st}^\varphi=\Delta t_{\rm int}+\Delta t^{\rm \mathcal{DM}}_{\rm 2nd}+\Delta t^\varphi_{\rm 2nd}.
\label{ddis4}
\end{equation}
Adopting the conservative assumption that the combined intrinsic and second-order plasma contributions satisfy $\Delta t_{\rm int}+\Delta t^{\rm \mathcal{DM}}_{\rm 2nd}\ge 0$, the residual delay provides a direct upper bound on the scalar-induced contribution $\Delta t_{\rm res}\ge \Delta t^\varphi_{\rm 2nd}$. This requirement yields the constraint
\begin{equation}
m_{\gamma_k}\leq 4\pi \Big[\frac{8\Delta t_{\rm res}}{3d}\Big(\frac{1}{\nu_1^4}-\frac{1}{\nu_2^4}\Big)^{-1}\Big]^{1/4}.  
\label{ddis5}
\end{equation}
on the effective scalar-induced photon mass. Using Eq.~\ref{import}, this bound can be translated into an upper limit on the scalar-photon coupling,
\begin{equation}
g_{\varphi\gamma\gamma}\leq \Big(\frac{320\pi}{B_0^2 R^3\Omega^2}\Big)^{1/2}\Big[\frac{8\Delta t_{\rm res}}{3d}\Big(\frac{1}{\nu_1^4}-\frac{1}{\nu_2^4}\Big)^{-1}\Big]^{1/8},  
\label{ddis6}
\end{equation}
which is valid for scalar masses satisfying $m_\phi<\Omega$. In obtaining this bound, we adopt the conservative geometric choices $\sin(2\alpha)\simeq 1$ and $\sin(2\theta)\simeq 1$.

Finally, we note that the scalar-induced photon mass oscillates in time with angular frequency $\Omega$. This characteristic time dependence provides an additional discriminator to the scalar-induced propagation effect from conventional astrophysical or instrumental backgrounds, which are not phase-locked to the source rotation.

\section{Constraints on the scalar-electron and scalar-photon couplings}\label{constraint}

\begin{table}[h]
\centering
\begin{tabular}{ |c|c|c|c|c|c|c| }
 \hline
 \multicolumn{4}{|c|}{Compact star parameters} \\
 \hline
 \hspace{0.01cm} & Crab pulsar\hspace{0.01cm}&SGR 1806-20\hspace{0.01cm}&GRB 080905A\hspace{0.01cm}\\
 \hline
 $P$ &$33~\mathrm{ms}$ \cite{Lyne:2014qqa}&$7.468~\mathrm{s}$ \cite{Marsden:2001tw}& $9.80~\mathrm{ms}$ \cite{Rowlinson:2013ue}\\
 $\dot{P}$ &$4.20\times 10^{-13}~\mathrm{s~s^{-1}}$ \cite{Lyne:2014qqa}&$115.7\times 10^{-12}~\mathrm{s~s^{-1}}$ \cite{Marsden:2001tw}& $1.86\times 10^{-7}~\mathrm{s~s^{-1}}$\cite{Rowlinson:2013ue}\\
$B_0$ & $(6.9-8.5)\times 10^{12}~\mathrm{G}$ \cite{Khelashvili:2024sup} & $(8-25)\times 10^{14}~\mathrm{G}$ \cite{MEREGHETTI20111317} & $  (27.1-49.5)\times 10^{15}~\mathrm{G}$ \cite{Rowlinson:2013ue}\\
$R$ & $14~\mathrm{km}$ \cite{Khelashvili:2024sup} & $10~\mathrm{km}$ \cite{Marsden:2001tw}  &$10~\mathrm{km}$ \cite{Rowlinson:2013ue}\\
$\alpha$ & $70^\circ$ \cite{Kou:2015oma} & $70^\circ$ \cite{Salmonson:2006jk} & $23^\circ$ \cite{Rowlinson:2010jb} \\
$dE/dt$ & $(4.49-4.51)\times 10^{38}~\mathrm{erg/s}$\cite{Khelashvili:2024sup}& $(0.5-1.4)\times 10^{36}~\mathrm{erg/s}$ \cite{MEREGHETTI20111317} & $(0.7-3.8)\times 10^{48}~\mathrm{erg/s}$ \cite{Rowlinson:2013ue} \\
\hline
\end{tabular}
\caption{\label{table4} Characteristic parameters for the candidate compact stars}
\end{table}

We analyze three compact-star systems--the Crab pulsar \cite{Lyne1993,Bejger:2002ty,Manchester:2004bp}, SGR 1806-20 \cite{sp2010,Younes:2015hsa}, and GRB 080905A \cite{Rowlinson:2010jb,Rowlinson:2013ue} to constrain the scalar-electron and scalar-photon couplings using various observations such as spin-down luminosity, photon redshift, surface magnetic field, and residual time delay. These sources are selected because their astrophysical parameters (magnetic field strength, rotation frequency, and emission characteristics) enable enhanced sensitivity to scalar-induced effects and because their sky locations have yielded high-quality observational data with comparatively smaller uncertainties. Together, these features make them among the most powerful astrophysical laboratories for placing competitive bounds on feebly coupled ultralight scalars.

Spatially varying, long-range ultralight scalar fields that couple to electrons and photons can induce deviations in the force law between macroscopic bodies, leading to apparent violations of Newtonian gravity and potential departures from Einstein's equivalence principle (EEP). Such effects are tightly constrained by high-precision tests of gravity, including torsion-balance measurements in the E\"ot-Wash experiments \cite{Hees:2018fpg,Adelberger:2003zx,Fischbach:1996eq,KONOPLIV2011401} and differential-acceleration measurements in the MICROSCOPE \cite{Berge:2017ovy} satellite mission. Independent constraints on scalar couplings also arise from stellar energy-loss arguments in astrophysical systems such as white dwarfs, red giants, where scalar emission can alter the brightness and surface temperature, enabling bounds on both scalar-electron and scalar-photon interactions. Importantly, all of the above probes constrain generic light scalar degrees of freedom and do not require the scalar to constitute the DM.

If, instead, a time-dependent scalar field forms a coherently oscillating ultralight DM background, additional sensitivity is gained through precision atomic, molecular, and optical measurements. In this regime, scalar couplings manifest as time-dependent variations in fundamental parameters, including the electron mass $m_e$ and the fine-structure constant $\alpha$, which can be probed using spectroscopic transitions and atomic clocks. A summary of such existing bounds can be found in \cite{AxionLimits}. Among all current constraints, the MICROSCOPE experiment \cite{Berge:2017ovy} provides the leading limit in the ultralight mass regime, delivering the most stringent test of EEP violation induced by scalar couplings to electrons and photons.

\subsection{Constraining scalar-electron coupling from spin-down luminosity}
Pulsars act as rapidly rotating, strongly magnetized NSs whose magnetic dipole axis is generically misaligned with the spin axis by an angle $\alpha$. This misalignment leads to EM dipole radiation, which carries away angular momentum and drains rotational kinetic energy of the stars. The rotational energy reservoir $E_{\rm rot}=\tfrac{1}{2}I\Omega^2$ therefore decreases as the spin slows-down, implying $\dot{\Omega}<0$, a growing spin period $P$, and a decreasing angular frequency $\Omega=2\pi/P$. The corresponding energy-loss rate can be expressed in terms of the measurable spin-down observable $\dot{P}$ as
\begin{equation}
\dot{E}_{\rm rot}=I\Omega\dot{\Omega}=I\frac{(2\pi)^2}{P^3}\dot{P}.
\label{erot}
\end{equation}
If the dominant spin-down channel is magnetic dipole radiation, the luminosity of which is
\begin{equation}
\dot{E}_{\rm dipole}=\frac{2}{3}(B_0 R^3 \sin\alpha)^2\Omega^4,
\label{edipole}
\end{equation}
then equating $\dot{E}_{\rm rot}=\dot{E}_{\rm dipole}$ yields the standard estimate for the surface magnetic field,
\begin{equation}
B_0\sin\alpha=\left(\frac{3I}{8\pi^2 R^6}\right)^{1/2}(P\dot{P})^{1/2}.
\label{compare}
\end{equation}

In addition to EM dipole radiation, GW emission can in principle contribute to the pulsar spin-down; however, for realistic NS ellipticities, the GW luminosity is several orders of magnitude smaller than the magnetic dipole loss and can be safely neglected. 

A quadrupole scalar field sourced by the time-dependent component of the magnetospheric GJ electron charge density provides an additional spin-down channel (Eq.~\ref{eq:13}), with an energy-loss rate that can be mapped onto the timing observable $\dot{P}$. Because the scalar charge distribution carries a $l=2$ structure, the resulting scalar radiation is also quadrupolar in the far zone. 

In the limit where the magnetic-dipole axis aligns with the rotation axis $\alpha \to 0$, the dipole moment becomes time-independent in the co-rotating frame, shutting off EM radiation and implying vanishing radiative spin-down power. Coherent scalar emission further requires that the scalar Compton frequency do not exceed the stellar rotation frequency, i.e., $m_\varphi < \Omega$; equivalently, scalar radiation is kinematically allowed only for $m_\varphi < \Omega$, and becomes ineffective once $m_\varphi > \Omega$. For ultralight scalars satisfying $m_\varphi<\Omega$, the induced quadrupole emission can lead to a monotonic increase of the spin period $P$, offering a principled avenue to test subdominant spin-down contributions from new light degrees of freedom beyond standard spin-down mechanisms.

To derive bounds on the scalar-electron coupling $g_e$, we compare the measured rotational energy-loss rate $\dot{E}_{\rm rot}$ (Eq.~\ref{erot}) with the expected loss from standard magnetic dipole radiation $\dot{E}_{\rm dipole}$ (Eq.~\ref{edipole}). Any extra contribution from scalar radiation $\dot{E}_\varphi$ (Eq.~\ref{eq:13}) must not exceed what is allowed by observations. We therefore impose
\begin{equation}
\dot{E}_{\rm rot} \ge \dot{E}_{\rm dipole} + \dot{E}_\varphi,
\end{equation}
and extract a bound on $g_e$ by requiring the scalar emission $\dot{E}_\varphi$ to lie within the $1\sigma$ observational uncertainty on $\dot{E}_{\rm dipole}$. This ensures that new scalar radiation does not overshoot the level already permitted by the precision of pulsar spin-down measurements.

Using the Crab pulsar as a benchmark, the scalar-radiation power can be expressed as
\begin{equation}
P_\Omega=0.01\times 10^{38}~\mathrm{erg/s}\Big(\frac{g_e}{0.05}\Big)^2\Big(\frac{B_0}{8.5\times 10^{12}~\mathrm{G}}\Big)^2\Big(\frac{R}{14~\mathrm{km}}\Big)^6 \Big(\frac{33~\mathrm{ms}}{P}\Big)^4,
\label{result1}
\end{equation}
where the limit is valid for the mass of the scalar $m_\varphi\lesssim 1.2\times 10^{-13}~\mathrm{eV}$, constrained by the spin angular frequency of the Crab pulsar, and we conservatively assume $\sin^2\alpha\sim1$. Considering that scalar radiation contributes to the $1\sigma$ measurement uncertainty of the observed spin-down luminosity, we obtain the bound on the scalar-electron coupling for the Crab pulsar as
\begin{equation}
g_e\lesssim 0.07.
\end{equation}
The resulting scalar-induced energy loss is likewise negligible for the other compact-star candidates considered in this paper.

\subsection{Constraining scalar-photon coupling from redshift measurement}

\begin{figure}
\includegraphics[width=14cm]{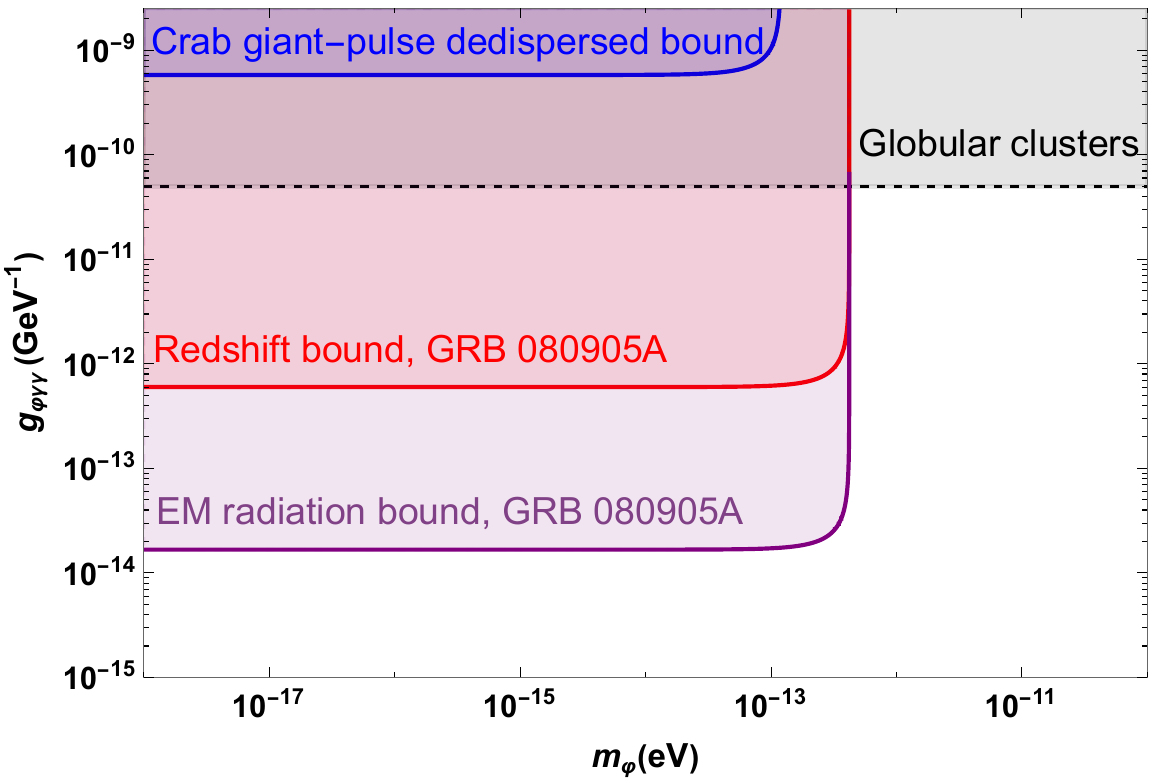}
\caption{Constraints on the scalar-photon coupling $g_{\varphi\gamma\gamma}$ derived from the frequency-dependent redshift measurement of GRB 080905A (red curve), its magnetic field estimation (purple curve) and from the residual time delay of dedispersed giant pulses from the Crab pulsar (blue curve). These bounds are compared with the existing astrophysical constraint from globular clusters (black dashed curve). The shaded regions denote the parameter space excluded by each probe.
}
\label{plot1}
\end{figure}

The frequency dependence of the redshift implied by Eq.~\ref{redshift-master} indicates that if multiple spectral lines from a magnetar can be observed corresponding to different photon frequencies and the redshift of the host galaxy is independently measured to provide a common normalization, it is possible to isolate the differential redshift contribution $\delta z(\omega)$. This would in turn allow an extraction of the scalar-photon coupling $g_{\varphi\gamma\gamma}$. To estimate the expected magnitude of this effect, we adopt GRB 080905A \cite{Rowlinson:2013ue,Ricci:2020oyt}, believed to originate from a magnetar, as a benchmark source. The apparent redshift, equivalently expressed as the fractional shift in the photon wavenumber, follows from Eq.~\ref{redshift-master} and can be written as
\begin{equation}
\begin{split}
\delta z\sim 10^{-4} \Big(\frac{g_{\varphi\gamma\gamma}}{4.5\times 10^{-13}~\mathrm{GeV}^{-1}}\Big)^4 \Big(\frac{2.1~\mathrm{GHz}}{\omega}\Big)^2\Big(\frac{B_0}{49.5\times 10^{15}~\mathrm{G}}\Big)^4\Big(\frac{R}{10~\mathrm{km}}\Big)^6\Big(\frac{9.80~\mathrm{ms}}{P}\Big)^4,   
\end{split}
\label{f1}
\end{equation}
where the limit is valid for the mass of the scalar $m_\varphi\lesssim 4.2\times 10^{-13}~\mathrm{eV}$, constrained by the spin angular frequency of the magnetar. We also chose conservative values $\sin^2{2\alpha}\sim1, ~\sin^2{2\theta}\sim1$. If redshift measurements can reach a precision at the level of $\sim10^{-4}$, this method would be sensitive to scalar-photon couplings as small as $g_{\varphi\gamma\gamma}\sim 4.5\times 10^{-13}~\mathrm{GeV}^{-1}$. Requiring the scalar-induced contribution to the photon redshift to remain within the $1\sigma$ observational uncertainty of the host-galaxy redshift for the GRB 080905A event, $\delta z \simeq 3\times10^{-4}$ \cite{Rowlinson:2010jb}, we derive an upper bound on the scalar-photon coupling of
\begin{equation}
g_{\varphi\gamma\gamma}\lesssim 6\times10^{-13}~\mathrm{GeV^{-1}}.
\end{equation}

There is also a constant (DC) term in Eq.~\ref{redshift-master}, which we neglect as it is challenging to isolate its effects from any static astrophysical backgrounds. The resulting constraint is shown by the red curve in FIG.~\ref{plot1}, with the shaded region indicating the excluded parameter space. This bound improves upon the existing globular cluster limit \cite{Dolan:2022kul} by approximately two orders of magnitude. Its explicit time-oscillating behavior with modulation frequency $2\Omega$ breaks any possible degeneracies with other astrophysical effects. The polar angle dependence on $\delta z(t)$ serves as an additional discriminator from non-scalar induced backgrounds. The induced frequency-dependent shift is enhanced for compact stars with stronger surface magnetic fields, larger radii, and emission detectable at lower photon frequencies, making magnetars particularly promising targets for such searches. Improved constraints are expected from differential, phase-resolved, and multi-frequency redshift measurements that suppress astrophysical systematics and are sensitive to the scalar-induced frequency shift.

\subsection{Constraining scalar-photon coupling from surface magnetic field measurement}
The interaction of a photophilic scalar with the background EM fields of a magnetized compact star induces additional EM field components, which appear as corrections to the standard background fields. In Eq.~\ref{sem5}, we derive the expression for the scalar-induced magnetic field. Its magnitude can be directly compared with that of the background dipolar magnetic field by interpreting it as a modification of the effective surface magnetic field at $r=R$.

Accordingly, the surface magnetic field of the compact star, $B_0$ (Eq.~\ref{compare}), receives contributions from both the standard EM field and the scalar-induced magnetic component $B_\varphi$ given in Eq.~\ref{sem5}. To place bounds on the photophilic coupling $g_{\phi\gamma\gamma}$, we parametrize the total surface magnetic field as
\begin{equation}
B_{\rm surf}^2 \simeq B_0^2 + B_\varphi^2.
\end{equation}
We then impose the conservative requirement that the scalar-induced contribution $B_\varphi$ does not exceed the $1\sigma$ observational uncertainty in the measured surface magnetic field $B_0$.

The strongest constraint on $g_{\phi\gamma\gamma}$ is obtained from the inferred surface magnetic field of GRB 080905A, yielding
\begin{equation}
g_{\varphi\gamma\gamma} \lesssim 1.7\times 10^{-14}~\mathrm{GeV^{-1}} ,
\end{equation}
for $m_\varphi\lesssim 4.2\times 10^{-13}~\mathrm{eV}$, as indicated by the purple shaded region in FIG.~\ref{plot1}. We also obtain relatively weaker bounds $g_{\varphi\gamma\gamma}\lesssim 3.1\times 10^{-10}~\mathrm{GeV^{-1}}$ for $m_\varphi\lesssim 1.2\times 10^{-13}~\mathrm{eV}$ from SGR 1806-20 and $g_{\varphi\gamma\gamma}\lesssim 1.1\times 10^{-10}~\mathrm{GeV^{-1}}$ for $m_\varphi\lesssim 1.2\times 10^{-13}~\mathrm{eV}$ from the Crab pulsar. The scalar-induced magnetic field decreases more rapidly with radial distance than the background dipolar field. The scalars with masses $m_\varphi \lesssim \Omega$ can contribute appreciably to the surface magnetic field. Moreover, the scalar-induced field is phase-locked to the stellar rotation frequency $\Omega$ and exhibits a characteristic polar-angle dependence. These distinctive features provide additional handles to discriminate the signal from conventional astrophysical backgrounds.
\subsection{Constraining scalar-photon coupling from residual time delay}
Pulsars are known to emit giant radio pulses with extremely large peak flux densities, reaching values in excess of $\mathrm{MJy}$, and with remarkably short durations ranging from microseconds down to nanoseconds. As reported in \cite{Hankins_2007}, a giant pulse from the Crab pulsar was observed at a central frequency of $9.25~\mathrm{GHz}$ with a bandwidth of $2.2~\mathrm{GHz}$, exhibiting a peak flux exceeding $2~\mathrm{MJy}$. The most intense nanoshot within this pulse had a dedispersed duration of $\Delta t_{\rm res}\lesssim 0.4~\mathrm{ns}$. Motivated by this observation, we take $\nu_1=8.15~\mathrm{GHz}$ and $\nu_2=10.35~\mathrm{GHz}$ as representative frequencies and adopt a distance of $d=2~\mathrm{kpc}$ for the Crab pulsar.

Using these benchmark values, we obtain the following upper bound on the scalar-photon coupling from Eq.~\ref{ddis6} as,
\begin{equation}
g_{\varphi\gamma\gamma}\lesssim 5.8\times 10^{-10}~\mathrm{GeV^{-1}},    
\end{equation}
which applies to scalar masses satisfying $m_\varphi\lesssim \Omega \simeq 1.2\times10^{-13}~\mathrm{eV}$. The corresponding constraint is shown by the blue curve in FIG.~\ref{plot1}, with the shaded region denoting the excluded parameter space. This bound is approximately one order of magnitude weaker than the existing globular cluster limit \cite{Dolan:2022kul}. Owing to its explicit dependence on the polar angle and the fact that the modulation frequency is locked to the stellar rotation frequency $\Omega$, the scalar-induced contribution to the residual time delay can also be distinguished from conventional $\mathcal{DM}$.

A more conservative estimate can be obtained by accounting for uncertainties in the $\mathcal{DM}$. Using the $\mathcal{DM}$ uncertainty of $10^{-5}~\mathrm{pc~cm^{-3}}$ quoted in \cite{Hankins_2007}, corresponding to a residual time delay of approximately $0.24~\mathrm{ns}$, the effective dedispersed duration increases to $\Delta t_{\rm res}\simeq 0.64~\mathrm{ns}$. Interpreting this value as an upper limit on the scalar-induced second-order delay $\Delta t^\varphi_{\rm 2nd}$ slightly weakens the bound, but does not alter its order of magnitude. Consequently, nanosecond-scale giant pulses from the Crab pulsar provide a robust and competitive probe of scalar-photon interactions in the ultralight mass regime. 

The constraints on the scalar-photon coupling can be further strengthened by considering alternative astrophysical sources and future observations. In particular, the detection of highly dedispersed, short-duration radio pulses, such as those from fast radio bursts (FRBs), offers a promising avenue, as FRBs have already provided some of the most stringent limits on the photon mass \cite{Chang:2022qct,Wei:2020wtf,Wang:2021nrl}. Similar improvements may also be achieved by targeting other strongly magnetized and rapidly rotating systems, including magnetars, where enhanced magnetic fields and high-precision timing can substantially increase sensitivity to scalar-induced propagation effects.

\section{Conclusions and discussions}\label{conclusion}

Compact objects such as pulsars and magnetars provide powerful astrophysical laboratories for probing ultralight scalar fields that couple to electrons or photons and may contribute to the DM energy density of the Universe. In this paper, we consider ultralight electrophilic scalars coupled to the GJ electron charge density in a corotating magnetosphere and study their impact on pulsar spin evolution. The coupling to the time-dependent quadrupolar component of the GJ charge density leads to scalar radiation from the magnetosphere, which acts as an additional channel for rotational energy loss and contributes to the secular increase of the spin period.

While scalar radiation sourced by GJ charge density from magnetized NSs has previously been explored under simplified assumptions \cite{Poddar:2024thb}, our analysis in this paper treats the NS as a rotating dipolar magnet in a more general framework. For the compact-star systems considered here, however, we find that the scalar-induced spin-down luminosity is many orders of magnitude smaller than the standard EM dipole contribution. Using Crab pulsar timing data, we obtain an upper bound on the scalar-electron coupling of $g_e \lesssim 0.07$ for scalar masses $m_\varphi \lesssim \Omega\sim1.2\times10^{-13}~\mathrm{eV}$, which is already excluded by existing laboratory, astrophysical, and fifth-force constraints. We derive the bound by requiring that the scalar-induced contribution to the pulsar spin-down luminosity does not exceed the $1\sigma$ observational uncertainty of the measured spin-down power. We further verify that other magnetar systems do not lead to a significant improvement of this bound.

The time-independent component of the scalar-induced GJ charge density sources a static, long-range quadrupolar scalar field with Yukawa suppression outside the magnetosphere. If such a field mediates an interaction between two magnetized compact stars, the resulting scalar force is modulated by a quadrupolar angular factor. This contribution is parametrically subleading compared to the Newtonian gravitational force and therefore has a negligible impact on the orbital dynamics of compact binaries.

Ultralight scalar fields are sourced by the time-dependent EM fields of magnetized compact stars, giving rise to a long-range, oscillatory quadrupolar scalar profile with a $1/r$ fall-off outside the star. As pulsar photons propagate through this time-dependent scalar background, their dispersion relation is modified, leading to a subluminal group velocity that can be interpreted due to an effective, scalar-induced photon mass. This modification results in a frequency-dependent redshift that oscillates in time with angular frequency $2\Omega$ and is present only for scalar masses satisfying $m_\varphi\lesssim\Omega$.

The characteristic frequency dependence of the induced redshift implies that if multiple spectral features from a magnetized star can be observed and the redshift of the host galaxy is independently measured to provide a common normalization, the differential redshift contribution $\delta z(\omega)$ can be isolated. Requiring the scalar-induced effect to remain within the $1\sigma$ uncertainty of the host-galaxy redshift measurement yields an upper bound on the scalar-photon coupling of $g_{\varphi\gamma\gamma}\lesssim 6\times10^{-13}~\mathrm{GeV^{-1}}$ for $m_\varphi\lesssim \Omega\sim 4.2\times 10^{-13}~\mathrm{eV}$ from the observation of GRB 080905A. While related constraints were previously obtained for photon propagation in static, monopolar scalar backgrounds \cite{Poddar:2025oew}, in this work, we study photon propagation through a time dependent scalar field background. This oscillatory behavior provides a distinctive signature that helps to disentangle scalar-induced redshift effects from conventional astrophysical backgrounds.

The induced fractional frequency shift maps directly onto pulsar timing residuals, defined as the difference between observed and model-predicted time of arrivals. The signal can therefore be searched for using coherent analyses of the timing residuals, employing matched-filter techniques or sinusoidal templates at the characteristic frequency $2\Omega$. This strategy enables sensitivity to extremely small, phase-coherent periodic effects and highlights the potential of precision pulsar timing as a probe of ultralight scalar interactions.

The interaction of a photophilic scalar with the vacuum EM fields modifies the background EM fields and, consequently, can alter the EM radiation emitted by a pulsar through corrections to the inferred surface magnetic field. Related effects were also investigated in \cite{Poddar:2025oew}; however, in that case the modification arises from photon propagation through a static scalar-field background. The scalar-induced magnetic field decreases more rapidly with radial distance than the standard dipolar field. Requiring that the scalar-induced contribution to the surface magnetic field remains within the $1\sigma$ observational uncertainty leads to a stringent constraint on the scalar-photon coupling, $g_{\varphi\gamma\gamma}\lesssim 1.7\times10^{-14}~\mathrm{GeV^{-1}}$, 
for scalar masses $m_\varphi\lesssim 4.2\times10^{-13}~\mathrm{eV}$. Furthermore, the scalar-induced magnetic field is phase-locked to the stellar rotation frequency $\Omega$ and exhibits a characteristic polar-angle dependence. These distinctive features provide additional discriminants that allow the signal to be separated from conventional astrophysical backgrounds.

The leading correction to the photon dispersion relation induced by the scalar field appears at $\mathcal{O}(\nu^{-2})$, sharing the same frequency dependence as the standard plasma $\mathcal{DM}$. If neither the time dependence of the signal nor polarization measurements alone are sufficient to lift the degeneracy associated with the $1/\nu^{2}$ frequency scaling, the degeneracy can be removed through de-dispersion techniques, which eliminate all $1/\nu^{2}$ contributions and allow higher-order terms, in particular those scaling as $\nu^{-4}$, to be probed in the residual time-delay signal. Existing measurements of dedispersed nanosecond-scale giant pulses from the Crab pulsar already enable such an analysis and lead to an upper bound on the scalar-photon coupling of $g_{\varphi\gamma\gamma}\lesssim 5.8\times10^{-10}~\mathrm{GeV^{-1}}$ for scalar masses $m_\varphi\lesssim \Omega\sim 1.2\times10^{-13}~\mathrm{eV}$. Observations of FRBs and other magnetar systems with larger magnetic fields could result in an even stronger bound on $g_{\varphi\gamma\gamma}$ from the residual time delay measurements. These results demonstrate that precision timing of short-duration radio bursts provides a viable and complementary probe of scalar-induced modifications to photon propagation.

In summary, the bounds we obtain on the scalar-electron coupling are not competitive with existing limits inferred from pulsar spin-down luminosities, reaffirming that scalar radiation contributes negligibly to the observed spin-down of pulsars. By contrast, we derive comparatively strong constraints on the scalar-photon coupling using redshift measurements, EM radiation, and dedispersed pulse-delay observations, which are in some cases stronger than the current astrophysical bounds. In particular, the limits on the scalar-photon coupling inferred from redshift measurements and EM emission from GRB~080905A surpass those derived from globular cluster observations by two to three orders of magnitude, respectively. These are the strongest astrophysical limits we obtain till date. Although the bounds remain weaker than constraints obtained from laboratory fifth-force experiments and the MICROSCOPE mission. Notably, the bounds presented here are independent of whether the scalar field constitutes a DM component of the Universe.

Looking ahead, these constraints can be significantly improved by targeting lower photon frequencies, where the sensitivity to scalar-induced propagation effects is enhanced. Radio facilities operating in the tens-of-MHz range, such as LOFAR \cite{Geyer:2017lbi,vanHaarlem2013} and the SKA \cite{Huege:2016jvc}, are therefore particularly promising. In addition, other highly magnetized astrophysical sources, especially FRBs, offer strong prospects for further improvement \cite{Shao:2017tuu,Wei:2020wtf,Xing:2019geq,Wang:2021nrl}. Observations of FRBs associated with large magnetic fields, for example through CHIME \cite{CHIMEFRB:2025crt} in combination with low-frequency radio measurements, could strengthen the bounds on scalar-photon couplings by several orders of magnitude. We will consider such cases in a separate publication.

\section*{Acknowledgments} 
This article is based on the work from COST Actions COSMIC WISPers CA21106 and BridgeQG CA23130, supported by COST (European Cooperation in Science and Technology).

\bibliographystyle{utphys}
\bibliography{reference}
\end{document}